# Fiber-induced nonlinearity compensation in coherent optical systems by affinity propagation soft-clustering

Elias Giacoumidis[1], Jinlong Wei[2], Ivan Aldaya[3], Christian Sanchez[4], Hichem Mrabet[5], & Liam P. Barry[1]

[1]Dublin City University, Radio and Optical Laboratory, Glasnevin 9, Dublin, Ireland.

[2]Huawei Technologies Düsseldorf GmbH, European Research Center, Riesstrasse 25, 80992 München, Germany.

[3]Sao Paolo University, São João da Boa Vista, Sao Paolo, Brazil.

[4]Aston University, Aston Institute of Photonic Technologies, Aston Triangle, Birmingham, England, UK.

[5]Saudi Electronic University, College of computation and Informatics, IT Department, Saudi Arabia.

(Email: elias.giacoumidis@dcu.ie).

**There is a tremendous need to maximize signal capacity without sacrificing energy consumption and complexity in optical fiber communications due to the rise of network traffic demand and critical latency-aware services such as telemedicine and the Internet-of-Things. In this work, the first system-agnostic and training-data-free fiber nonlinearity compensator (NLC) using affinity propagation (AP) clustering is experimentally demonstrated. It is shown that compared to linear equalization, AP can increase the signal quality-factor up to ~5 dB in 16-quadrature-amplitude-modulated coherent optical systems by effectively tackling deterministic and stochastic nonlinearities. AP outperforms benchmark clustering algorithms and complex deterministic schemes such as K-means, fuzzy-logic, digital-back propagation and Volterra-based NLC, offering a power margin extension of up to 4 dB.**

There has been intensifying focus on minimizing energy-efficiency while increasing transmission speed in optical fiber communications for applications such as warehouse-scale data centers[1], access, metropolitan networks[2] (including 4G/5G[3]) and even long-haul networks due to the rise of the network traffic demand. Cisco predictions announced an annual increasing of 27% in the IP traffic mainly due to the popularity of the high-bandwidth video content[4], and metro network traffic increased twofold in comparison with backbone traffic in 2017[5]. Moreover, the rapid growth of real-time machine-to-machine communications such as remote medicine, financial trading and the Internet-of-Things, mandates the realization of latency-aware optical networks[6]. To increase data rates, one of the core difficulties is the optical Kerr effect that arises at high signal powers causing a variation in index of refraction which is proportional to the local irradiance of the light[7] and is responsible for nonlinear optical effects such as self-phase modulation, cross-phase modulation (XPM), and four-wave mixing (FWM) that generates a fourth idler photon[8]. The optical Kerr effect is attributed to the so-called nonlinear Shannon capacity limit[9] which sets an upper bound on the achievable data rate in optical fiber communications when using traditional linear transmission techniques. A conceptual diagram is depicted in Fig. 1, displaying the evolution of the quality of an optical signal over an optical fiber in frequency and time domain. It is shown that the signal is distorted over longer length of optical fiber, where especially after optical amplification stochastic nonlinearities play a crucial role. In the inset of Fig. 1, simulated quaternary phase-shift keying (QPSK) received constellation diagrams for 20 Gbit/sec coherent optical orthogonal frequency division multiplexing (CO-OFDM) are shown (at a fixed launched power) when linear equalization is only implemented. The constellation QPSK diagram is clearly distorted for longer length of optical fiber due to the interaction of fiber-induced nonlinearity with optical amplification noise.

On the other hand, there have been extensive efforts in attempting to surpass the nonlinear Shannon limit through several fiber nonlinearity compensation techniques and nonlinear transmission schemes[10–15] that compensate the deterministic Kerr-induced nonlinear effects. Albeit the Kerr-mediate nonlinear process is deterministic, the frequency uncertainty of many independent wavelength channels is transformed into time uncertainty through fiber transmission by chromatic dispersion, making the nonlinear interaction appear random[10]. This problem has been partially tackled by a combination of optical frequency combs (OFCs) and digital back-propagation (DBP)[10]. However, OFCs are impractical in long-haul coherent optical networks while DBP is extremely complex and energy-ineffective for real-time signal processing[10,12].

Moreover, long-haul optical networks involve several optical amplifiers (Erbium-doped fiber amplifiers, EDFAs) to compensate fiber losses, causing stochastic parametric noise amplification (PNA)[11] by means of the interplay between amplified spontaneous emission (ASE) noise (acting as Gaussian noise) and fiber nonlinearity. PNA is particularly debilitating for long-haul networks that utilize cascaded EDFAs, where the lower optical signal-to-noise ratio would prompt interest in nonlinearity compensation, since it scales at least quadratically with the system length[9]. Whilst the need for competitive edge ensures commercial interest in practically implementable form of nonlinearity compensation, it is difficult to imagine that a 50% increase in capacity would postpone an upcoming "capacity crunch" for the long-term[9]. Furthermore, albeit the Kerr-induced nonlinear process is deterministic, in multicarrier schemes like CO-OFDM[15,16] the resulting nonlinear interaction between subcarriers becomes very complicated appearing random due to its high peak-to-average power ratio (PAPR)[16]. Recently, unsupervised and supervised machine learning such as K-means clustering[17–19] and artificial neural network regression[3] have been introduced in optical communications to combat stochastic sources of noise, performing blind and non-blind nonlinear equalization (NLE), respectively. However, their performance benefit in both long-haul single-carrier and CO-OFDM have been limited due to their inability to effectively compensate the strong nonlinear phase noise (stochastic & deterministic) in higher-order modulation formats.

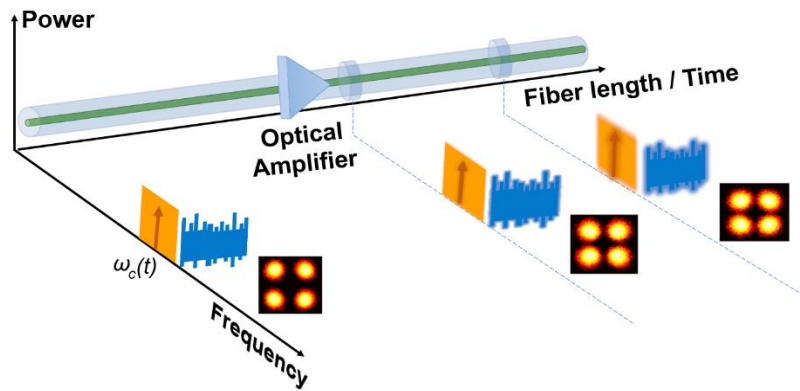

**Figure 1 | Conceptual diagram related to the impact of stochastic nonlinearities on optical signals.** Frequency and time domain signal quality evolution over an optical fiber that includes optical amplification. Inset: Three cases of received quaternary phase-shift keying (QPSK) constellation diagrams for coherent optical OFDM (CO-OFDM) at 20 Gbit/sec (simulated). $\omega_c(t)$: Optical carrier angular frequency at a time *t*.

In this work, we experimentally demonstrate the first affinity propagation (AP) clustering based completely blind-NLE that is system-agnostic without requiring training-data for a ~18.2 Gbit/sec QPSK middle-channel wavelength division multiplexing (WDM)-CO-OFDM and a ~40 Gbit/sec 16-quadrature amplitude modulated (16-QAM) single-channel CO-OFDM at 32000 km and 2000 km of standard single-mode fiber (SSMF) transmission, respectively. Results show that AP outperforms benchmark machine learning clustering algorithms for blind-NLE such as fuzzy-logic c-means (FL)[17] and K-means by up to > 1 and 2 dB in signal quality (Q)-factor, respectively. We also show that AP can improve the Q-factor by up to 3 and 5 dB compared to the widely-adopted deterministic approaches of full-step DBP (FS-DBP) and Volterra-based NLE[20], respectively, by providing a significant amount of stochastic nonlinear noise reduction mainly on middle OFDM subcarriers.

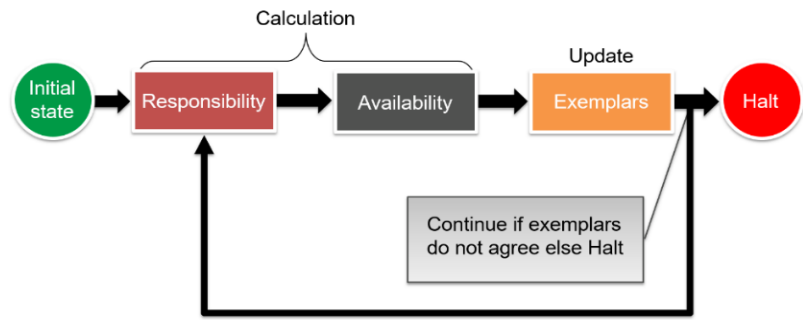

**Figure 2 | Affinity propagation (AP) algorithm.** AP steps that include the responsibility, availability and exemplars matrices.

**Nonlinearity cancellation using affinity propagation**

In machine learning clustering, defined data learn a set of centers such that the sum of squared errors between data points and their nearest centers is small[21]. When the centers are selected from actual data points, they are called "exemplars". The well-known K-means clustering begins with an initial set of randomly selected exemplars and iteratively refines this set so as to decrease the sum of squared errors[17]. In AP however, every symbol is a potential exemplar and by viewing each symbol as a node that recursively transmits real-valued messages (separately for amplitude and phase) along the edges of the NLE network until a good set of exemplars and corresponding clusters emerge. 'Messages' are updated by simple formulas that search for minima of an appropriately chosen energy function[21]. At any symbol in time, the magnitude of each message reflects the current affinity that 1 symbol has for selecting another symbol as its exemplar. Let $x_1$ through $x_n$ be a set of complex data (symbol), with no assumptions made about their internal structure, and let S be a function that quantifies the similarity between any 2 symbols, such that $S(x_i, x_J) > S(x_i, x_k)$ if $x_i$ is more similar to $x_J$ than to $x_k$. For this example, the negative squared distance of 2 symbols is used i.e. for points $x_i$ and $x_k$. The diagonal of S (i.e. S(i,i)) is particularly important, as it represents the input preference, meaning how likely a particular input is to become an exemplar. When this is set to the same value for all inputs, it controls how many classes the algorithm can produce. A value close to the minimum possible similarity produces fewer classes, however, a value close or larger to the maximum possible similarity, produces many classes (initialized to the median similarity of all pairs of inputs). AP proceeds by alternating 2 message passing steps to update the 'responsibility, R(i,k)' and 'availability, A(i,k)' matrices, where R quantifies how "well-suited" $x_k$ is to serve as the exemplar for $x_i$ compared to other candidate exemplars, while A shows how "appropriate" it would be for $x_i$ to pick $x_k$ as its exemplar, taking into account other points' preference. R and A, are initialized to zero and are viewed as log-probability tables and then AP is iteratively updated for R and A by the following expressions:

$$R(i, k) = s(i, k) - \max_{k' \neq k}\{a(i, k') + s(i, k')\} \quad (1)$$

$$A(i, k) = \min\left(0, r(k, k) + \sum_{i' \notin \{i,k\}} \max(0, r(i', k))\right)_{\text{for } i \neq k} \quad (2)$$

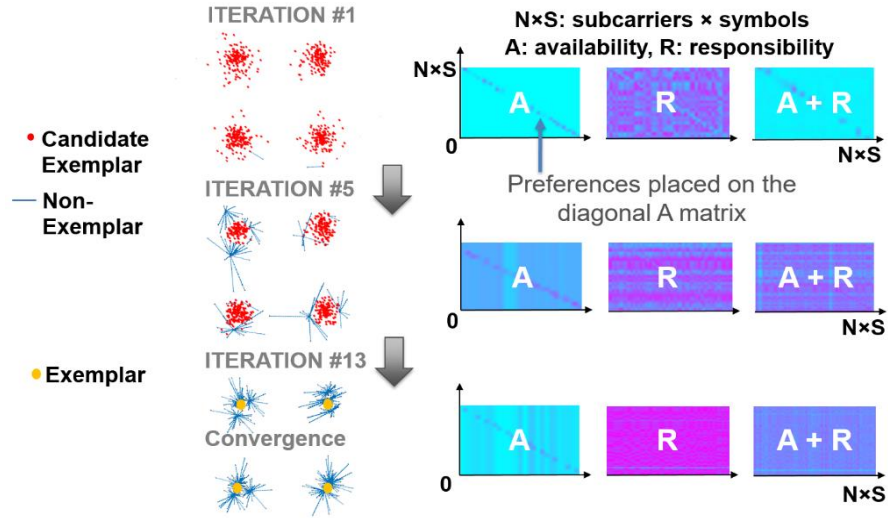

**Figure 3 | AP clustering example steps.** AP procedure for a QPSK CO-OFDM signal. 13 iterations are required for the AP algorithm to converge.

The exemplars are extracted from the final updated matrices where 'responsibility + availability' is positive[21]. In Fig. 2, the steps of the AP algorithm are presented, while Fig. 3 shows the aforementioned AP iterative result of R and A for a QPSK middle-channel in WDM CO-OFDM at 3200 km of SSMF transmission at an optimum launched optical power (LOP) per channel of −5 dBm. In Fig. 3, it is evident that 13 iterations are required for the system to convergence. Finally, it should be noted that while K-means and AP belong to the deterministic types of clustering, a probabilistic type also exist such as FL[17], permitting the symbols to fluctuate the data membership degree. In this work, we compare AP with FL and K-means based clustering which perform overlapping/soft and exclusive/hard and clustering, respectively. However, since AP also executes overlapping clustering, it is considered as the first overlapping deterministic clustering approach in optical fiber communications that can also potentially benefit wireless[22] telecommunications.

**Benchmark clustering and deterministic nonlinearity compensators**

K-means clustering is based on the Lloyd's algorithm[23], and is an iterative, data-partitioning algorithm that assigns *n* observations to exactly one of *K* clusters defined by centroids; where *K* is chosen before the algorithm starts. The algorithm proceeds as follows:

1. Choose *K* initial cluster centers (centroid);
2. Compute point-to-cluster-centroid distances of all observations to each centroid;
3. Assign each observation to the cluster with the closest centroid (Batch update);
4. Compute the average of the observations in each cluster to obtain *K* new centroid locations;
5. Repeat steps 2 through 4 until cluster assignments do not change, or the maximum number of iterations is reached.

FL is a probabilistic machine learning algorithm, permitting the symbols to fluctuate the data membership degree (MD) while being allocated into many clusters by minimizing the objective following function:

$$F_m = \sum_{i,j}^{N} \sum_{i=1}^{R} \sum_{j=1}^{L} \mu_{ij}^m \|t_i - c_j\|^2. \tag{3}$$

where N, R, L, and m, are the total number of OFDM subcarriers, symbols, clusters, and the "Fuzzy partition matrix exponent", respectively, the latter which regulates the 'degree' of clusters overlapping for m>1. Such overlapping is related to a fuzzy one, denoting the degree of boundaries' fuzziness between clusters[17]. Where $t_i$ is referred to the i$^{th}$ symbol, $c_j$ is the center of a j$^{th}$ cluster, and $\mu_{ij}$ refers to the MD of $t_i$ into j$^{th}$ cluster. FL is processed in five steps:

1. Enter the number of targeted clusters;
2. Initiate the cluster MD, $\mu_{ij}$;
3. Estimate the center per cluster by (4);
4. Update $\mu_{ij}$ using (5) and compute $F_m$;
5. Return and perform 2–4 steps until $F_m$ is converged at a desired threshold.

$$C_j = \sum_{i,j}^{N} \left( \sum_{i=1}^{R} \mu_{ij}^m X_i / \sum_{i=1}^{R} \mu_{ij}^m \right) \tag{4}$$

$$C_j = 1/\left(\sum_{i,j}^{N} \sum_{k=1}^{L} \|t_i - c_J\|/\|t_i - c_k\|\right)^{2/m-1} \qquad (5)$$

Finally, FS-DBP[12] and Volterra-based NLE[20] solve deterministically the inverse nonlinear Schrödinger equation that expresses the inverse parameters of the SSMF link. However, DBP uses the split-step Fourier method to compensate the chromatic dispersion and fiber nonlinearity, thus employing multiple fast Fourier transform (FFT) and inverse FFT (IFFT) blocks, respectively (i.e. 40 steps/span for FS-DBP). To relax computational effort, Volterra-based NLE has been introduced using Kernel filtering via the inverse Volterra-series transfer function. In Volterra-NLE, chromatic dispersion and nonlinearity are compensated per span (in contrast to DBP is independent from fiber length), and all spans are processed in parallel. Up to 2nd order Kernels have been used here to account for 2nd order chromatic dispersion, above which was not necessary (to maintain low complexity) as no performance improvement was observed[20]. It should be noted though, that it has been shown that Volterra-NLE is more complex than supervised and unsupervised machine learning algorithms[24].

**Experimental setup and results**

**Experiments using multi-carrier coherent optical signals**

The experimental setup is shown in Fig. 4 for both single-channel and WDM CO-OFDM at 2000 km and 3200 km of transmission, respectively, with procedures similar to[17]. In the digital signal processing (DSP) units, 400 OFDM symbols (20.48 ns length) were generated using a 512-point IFFT on 210 QPSK/16-QAM subcarriers. To eliminate inter-symbol-interference from linear effects, a cyclic prefix (CP) of 2% was included. For clustering, FS-DBP, Volterra-based NLE and without using NLE (i.e. linear equalization), the raw bit-rates were ~18.2 Gbit/sec (QPSK) and 40 Gbit/sec (16-QAM). The offline OFDM demodulator included timing synchronization, frequency offset compensation, channel estimation and equalization with the help of an initial training sequence, as well as IQ imbalance and chromatic dispersion compensation using an overlapped frequency domain equalizer (via the overlap-and-save method)[17]. For the WDM CO-OFDM system, a laser grid of 100 kHz-linewidth distributed feedback lasers (DFBs) on 100 GHz grid was used with the help of a polarization maintaining multiplexer (PMM), which were substituted in turn by a 100 kHz linewidth laser. The 100 kHz linewidth DFBs were located between 193.5–193.9 THz. Additional loading channels (10 GHz of bandwidth) were generated using an amplified spontaneous emission (ASE) source that were spectrally shaped using a wavelength selective switch (WSS). The 20 loading channels (100 GHz spacing from centre-to-centre channel) were spread symmetrically around the test wavelengths so that the total bandwidth of the transmitted signal was 2.5 THz. An IQ Mach-Zehnder modulator (IQ-MZM) was used to drive the complex OFDM data (2 channels × 25 GSamples/sec). A wide optical band-pass filter (BPF) was used to filter out-of-band ASE noise at the transmitter. The transmission path was an acousto-optic modulator (AOM) based re-circulating loop consisting of 4×100 km spans of SSMF. After propagation, the signal was filtered using a 4.2 nm flat topped filter and coherently detected.

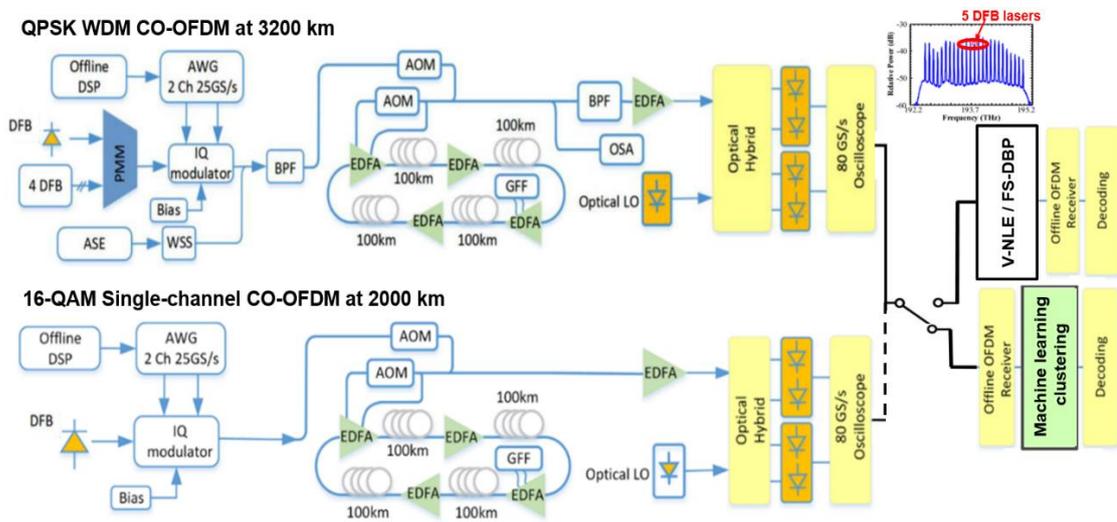

**Figure 4 | Experimental setup.** Single- and multi-channel CO-OFDM for 16-QAM (2000 km) and QPSK (3200 km), respectively, incorporating at the receiver digital signal processing (DSP) unit either machine learning clustering or deterministic nonlinearity compensators (Volterra-based nonlinear equalization, V-NLE, or full-step digital back-propagation, FS-DBP). Inset: Received optical spectrum for WDM CO-OFDM. AWG: arbitrary waveform generator; PMM: polarization maintaining multiplexer; WSS: wavelength selective switch; DFB: distributed feedback laser; AOM: acousto-optic modulator; GFF: gain-flatten filter, BPF: bandpass filter; LO: local oscillator; EDFA: Erbium-doped fiber amplifier; ASE: amplified spontaneous emission; OSA: optical spectrum analyser.

For the single-channel CO-OFDM similar but simpler setup was used using only an IQ-MZM at the transmitter, however, the transmission path at 1550.2 nm required a recirculating loop with 20×100 km spans of SSMF controlled by AOMs. For both systems, a traditional homodyne coherent optical receiver was used at the receiver part. Further details are included in Methods. The NLEs performances were assessed at the receiver-DSP unit by the bit-error-rate and the Q-factor ($=20\log 10[\sqrt{2}erfc^{-1}(2BER)]$) measurements averaging over 10 recorded traces (~$10^6$ bits) by error counting (hard-decision decoding, HDD).

**Performance comparison of affinity propagation with benchmark nonlinear equalizers**

In Fig. 5, the Q-factor against the LOP per channel is plotted for all adopted NLEs for QPSK (middle channel) WDM CO-OFDM, while in Fig. 6 the related received constellations diagrams are depicted for AP and K-means at a low LOP of −7 dBm where ASE noise dominates. As shown from Fig. 5, AP tackles more effectively nonlinearities compared to all algorithms under test, i.e. FL, K-means, FS-DBP and Volterra-based NLE (or else called for simplicity V-NLE). More specific, AP presents a performance benefit over the whole range of LOPs. At low powers, such as at −7 dBm in Fig. 6, AP outperforms the other clustering algorithms by tackling more successfully the ASE noise and some residual inter-channel crosstalk effects. This is corroborated in the simulation curves (with procedure described in Methods) for linear equalization (i.e. without using NLE), where a difference of < 0.5 in Q-factor is observed compared to the experimental case due to implementation penalty, which is mainly induced from electrical noise, and therefore, we assume we tackle in the largest proportion ASE noise and the residual inter-channel crosstalk effects at low powers. In the example of Fig. 6, we show that the ability of overlapping clustering has an advantage over the exclusive clustering of K-means. At optimum and high LOPs, the AP performance improvement can be explained by the fact that it tackles better the PNA effect and the accumulated intra- and inter-subcarrier FWM which appears random (due to CO-OFDM's high PAPR). Intra-channel nonlinearities are evident in Fig. 7, where the Q-factor is plotted at optimum −5 dBm of LOP for the middle OFDM subcarriers which suffers the most mainly from nonlinearities. More specific, at optimum LOP, AP provides up to > 1 dB of Q-factor enhancement on middle subcarriers compared to FS-DBP, confirming that in CO-OFDM the intra-channel nonlinearities are not effectively compensated by deterministic methods. It should be noted that the remarkable improvement of our nonlinear equalizer over FS-DBP in the WDM configuration is due to the fundamental inability of DBP to tackle inter-channel and inter-carrier nonlinear crosstalk effects without an accurate knowledge of the absolute position and the relative separation of both the channels and subcarrier frequencies[10]. We should note that the statistical nature of our proposed scheme can tackle both deterministic and stochastic nonlinear effects even without knowledge of the data. Moreover, since we consider a WDM and not an ultra-densed WDM CO-OFDM system, some of the inter-channel nonlinearities fall within the channel "guard-band".

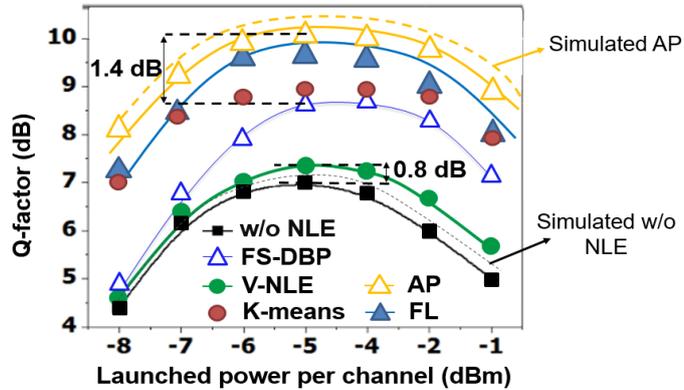

**Figure 5 | AP performance in WDM-QPSK.** Q-factor vs. launched optical power (LOP) per channel for QPSK WDM CO-OFDM at 3200 km of fiber transmission using AP, FL, K-means, FS-DBP, V-NLE and without (w/o) using NLE. The simulated AP and w/o NLE is also depicted.

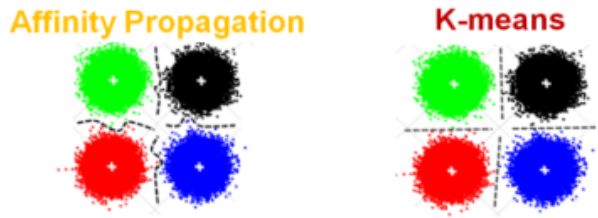

**Figure 6 | Received QPSK constellation diagrams.** Diagrams are at –7 dBm of LOP (per channel) in multi-channel CO-OFDM for AP and K-means. White crosses denote the centres of the clusters/constellation points.

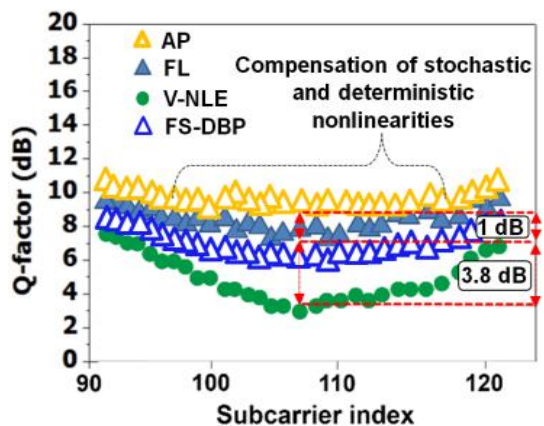

**Figure 7 | AP performance on middle subcarriers in WDM-QPK.** Middle subcarriers Q-factor distribution for multi-channel QPSK CO-OFDM at optimum –5 dBm of LOP per channel.

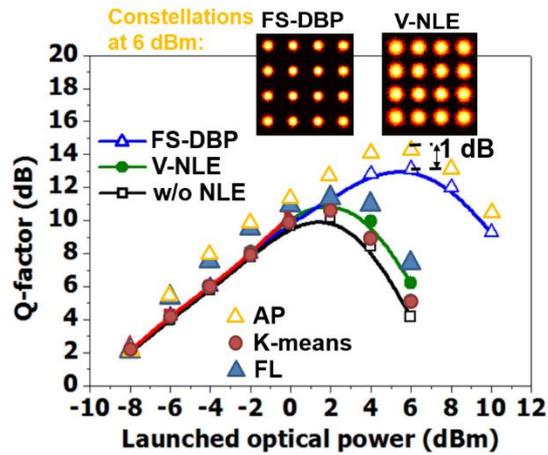

**Figure 8 | AP performance in single-channel 16-QAM.** Q-factor vs. LOP in single-channel 16-QAM CO-OFDM for machine learning and deterministic algorithms. Inset: Received 16-QAM constellation diagrams for FS-DBP and V-NLE at 6 dBm of LOP.

In Fig. 8, results are shown for single-channel 16-QAM CO-OFDM at 2000 km of fiber transmission, where the insets present the received constellation diagrams corresponding to the deterministic methods at optimum 6 dBm of LOP. As shown in Fig. 8, AP significantly outperforms linear equalization at optimum powers, offering ~5 dB of Q-factor enhancement. AP is also very effective for single-channel 16-QAM at high LOPs increasing the Q-factor by 1 dB over FS-DBP at optimum 6 dBm of LOP, while significantly outperforming the benchmark clustering algorithms and V-NLE. Compared to FL the proposed AP enables advanced soft clustering resulting in 4 dB extension of the required optimum LOP (increase in power margin) and up to 1 dB Q-factor improvement at low powers. However, the performance benefit over FS-DBP in 16-QAM is not as high as in WDM-QPSK due to the stronger nonlinear phase noise which affect the nonlinear decisions mainly on the outer clusters. On the other hand, the performance improvement of AP at low powers is not as significant as in QPSK since the circular-noisy-distributed outer clusters in 16-QAM are unpredictable distorted by Gaussian noise.

## Discussion

We experimentally demonstrated the first machine learning based blind nonlinearity compensator using affinity propagation clustering for fiber-optic communications beyond the nonlinear Shannon-limit. The proposed unsupervised (training-data-free) machine learning technique offered up to 4 dB extension of power margin in coherent optical OFDM by compensating nonlinear crosstalk effects. Affinity propagation outperformed benchmark clustering approaches such as K-means and fuzzy-logic c-means due to its ability for advanced soft-clustering (overlapping) of complex data. As a result, the proposed solution tackled more effectively nonlinearities (especially inter-subcarrier FWM on middle subcarriers), accumulated amplified spontaneous emission noise from concatenated optical amplifiers and its interplay with nonlinearity, i.e. the parametric noise amplification effect. It should be noted that since the digital-to-analogue/analogue-to-digital converters (DACs/ADCs) and transceiver noise was negligible as demonstrated by the small implementation penalty in the simulated curves of Fig. 5 (also ensuring an error vector magnitude of < 7% in 'optical back-to-back'), we assume that most of the performance benefit at low powers originates from the reduction of the optical amplification noise and the residual inter-channel crosstalk effects (for the WDM case). For high-order modulation formats such as 16-QAM, affinity propagation presented better capability of tackling nonlinearities in comparison to alternative machine learning clustering algorithms. For instance, compared to fuzzy-logic c-means, our proposed scheme for 16-QAM increased the system power margin by 4 dB. In comparison to the complex deterministic approaches of full-step digital back-propagation and Volterra-based nonlinear equalization, our solution enhanced the signal quality-factor by up to 3 dB. At low powers in the WDM case, a much-improved signal quality-factor was observed by affinity propagation compared to the full-step digital back-propagation due to its ability of tackling deterministic intra-channel nonlinearities. On the other hand, in the single-channel case a small performance difference between the two techniques was notified at low launched optical powers. This confirms the fact that in WDM the performance difference comes from other sources of noise besides amplified spontaneous emission and intra-channel nonlinearities, which is inevitably the remaining inter-channel nonlinearities, assuming that electrical noise is negligible and polarization-mode dispersion is solely compensated in long-haul systems by the OFDM's cyclic prefix length (2%).

Compared to other benchmark clustering approaches and supervised machine learning algorithms such as hierarchical[17] clustering, artificial neural networks[16,24,25] and support vector machines[26–30]; affinity propagation outperforms in single- and multi-channel coherent optical OFDM (up to 2 dB in quality-factor). This reveals that, to the best of our knowledge, the proposed scheme provides the highest performance (with the only exception when comparing to more sophisticated supervised three-dimensional deep neural networks[31]) in long-haul coherent optical multi-carrier schemes.

A pseudorandom unrepeated sequence was employed in this work with length of $2^{19}-1$ having a period of approximately $2^{19937}-1$ (Mersenne twister[32]). Compared to the work reported in[33] showing that when employing short pseudorandom sequences (with lengths of $2^7$ and $2^{15}$), supervised neural networks most likely will overestimate the system performance, our adopted pseudorandom sequence has a much longer period and since no training data are considered results were not overestimated. Our digital signal processing algorithm is transparent to single-carrier modulation techniques and other multi-carrier schemes such as Nyquist-WDM[34,35] and filter-bank[36,37] that have dominated modern coherent optical communications and it can seamlessly operate in low-cost direct-detected systems and dual-polarization signals by using a 2×2 multiple-input multiple-output (via the zero-forcing algorithm)[38] digital block of butterfly structure to unravel the polarization crosstalk. Due to experimental set-up limitations we did not demonstrate a dual-polarization system and hence is out of the scope of this work. However, in future experiments we are planning to address this by using two affinity propagation equalizers (for X- and Y-polarization). We will also consider higher-order modulation formats such as 64-QAM where it is anticipated a performance benefit but with slightly more complexity due to the higher number of clusters/constellation points. We envisage that digital nonlinear equalization using affinity propagation could potentially benefit 5G wireless[39,40], radio-over-fiber (RoF)[41] communications, visible light communications[42,43], as well as satellite transmission systems[44]. It worth mentioning that field-programmable gate-array (FPGA) based real-time implementation of the proposed algorithm has been launched through collaboration with Xilinx-Ireland[45] for RoF and long-haul coherent optical systems. Finally, we believe that the proposed solution could also benefit very-high spectrally-efficient modulation formats such as Fast-OFDM[46-50].

**Methods**
**Experimental setup**. In the transmitter, the single- and multi-channel CO-OFDM systems employed external cavity lasers (ECLs) of 100 KHz linewidth, being modulated using a dual-parallel IQ-MZM fed with 'offline' OFDM I-Q components. Single-polarization CO-OFDM was also considered for both single- and multi-channel configurations. The transmission path at 1550.2 nm was a recirculating loop consisting of 20×100 km (single-channel) and 32×100 km spans of Sterlite OH-LITE fiber (attenuation of 18.9–19.5 dB/100 km) controlled by acousto-optic modulator. The loop switch was located in the mid-stage of the 1st EDFA and a gain-flattening filter was placed in the mid-stage of the 3rd

EDFA for both configurations. For WDM CO-OFDM, four distributed feedback lasers and an ECL were employed on 100 GHz grid located between 193.5–193.9 THz. Using an ASE source, another 20 'dummy' WDM channels of 10 GHz bandwidth were generated. These channels covered 2.5 THz of bandwidth as depicted in the inset of Fig. 4. In the experiments, the baseband waveform samples were calculated offline based on a pseudo-random binary sequence of $2^{19}$-1. In the transmitter, an arbitrary waveform generator was used at a sampling rate (bandwidth) of 34 GHz (2 channels with 25 GSamples/sec for I and Q components) to generate a continuous baseband signal for both single-channel and WDM (middle-channel) CO-OFDM. At the receiver, for both single- and multi-channel systems, the optical receiver was constituted by a homodyne coherent detector with 100 kHz linewidth local oscillator laser (ECL-based). An oscilloscope was used at 80 GSamples/sec in real-time and the digital part was performed offline in Matlab®. 210 subcarriers were generated in a 512 inverse-FFT (IFFT) size using QPSK (WDM) and 16-QAM (single-channel) formats. Table I below presents the key experimental transceiver and transmission parameters.

**Table I. Key experimental transceiver and transmission parameters.**

| Parameters | Single-channel CO-OFDM | WDM-CO-OFDM |
|---|---|---|
| Modulation format | 16-QAM | QPSK |
| Raw signal bit-rate | ~46 Gbit/sec | ~20 Gbit/sec |
| Nominal signal bit-rate | ~40 Gbit/sec | ~18.2 Gbit/sec |
| Number of CO-OFDM channels | 1 | 20 (5 from lasers) |
| Total bandwidth (channel spacing) | 34 GHz (–) | 2.5 THz (10 GHz) |
| Transmission-reach | 2000 km [20×100 km spans] | 3200 km [32×100 km spans] |
| OFDM cyclic prefix length | 2 % | 2 % |
| ECL/DFB linewidth | 100 kHz | 100 kHz |
| Generated subcarriers (IFFT size) | 210 (512) | 210 (512) |

**Simulation setup**. The developed single-carrier simulated optical system was implemented in a Matlab/Virtual Photonics Inc. (VPI)-transmission-Maker® co-simulated environment (electrical domain in Matlab and optical components with standard single-mode fiber in VPI). The DACs/ADCs clipping ratio and quantization have been considered and set at optimum 13 dB and 10-bits (which does not limit OFDM signals for up to 256-QAM[38]). For the in-line optical amplification, EDFAs were adopted having with 5.5 dB of noise figure. The EDFA noise was also modelled as additive white-Gaussian noise. The optical fiber for single-polarization transmission was modelled using the pseudo-spectral split-step Fourier method which solves the nonlinear Schrödinger equation. The adopted standard single-mode fiber parameters in this work are the following: fiber nonlinear Kerr parameter, chromatic dispersion, chromatic dispersion-slope, fiber loss, and polarization-mode dispersion coefficient of 1.1 $W^{-1} km^{-1}$, 16 ps $nm^{-1} km^{-1}$, 0.06 ps $km^{-1}$ $(nm^2)^{-1}$, 0.2 dB $km^{-1}$ and 0.1 ps $(km^{0.5})^{-1}$, respectively. The signal bit-rate was set to the exact same value to the experimental CO-OFDM for both QPSK and 16-QAM formats. Single single-polarization was used, while the receiver linear equalizers performed similarly to the experiments a frequency domain equalizer and pilot-assisted frequency offset compensation and carrier phase estimation[38].

**Acknowledgements**

The work of E.G. was emanated from EU Horizon 2020 research and innovation programme under the Marie Skłodowska-Curie grant agreement No 713567 and in part by a research grant from Science Foundation Ireland (SFI) and is co-funded under the European Regional Development Fund under Grant Number 13/RC/2077. The work of J.W. was supported by the EU Horizon 2020 research and innovation programme under the Marie Skłodowska-Curie grant agreement No 623515. We thank Sterlite Techn. for letting us the OH-LITE optical fiber and Prof Andrew D. Ellis from Aston University for his help with the experiment setup. We would also like to thank for their contribution to this work: Dr Son T. Le (Aston University) now with Nokia Bell-Labs, Germany; and Dr Mary E. McCarthy (Aston University) now with Oclaro, UK.

**Author contributions**

E.G, J.W., I.A., C.S., H.M., and L.P.B. developed the basic idea for the generation of the affinity propagation-based machine learning equalizer for high-speed coherent optical communications. E.G. proposed the proof-of-concept experimental set-up. E.G. conducted the experiments at Aston University. All authors contributed to the writing of the manuscript.